\documentclass[
reprint, 
superscriptaddress, 
amsmath,
amssymb, 
floatfix,
longbibliography
]{revtex4-1}

\usepackage{graphicx}
\usepackage{xcolor}
\usepackage[colorlinks=true,citecolor=blue,linkcolor=magenta]{hyperref}
\usepackage{bm}

\begin{document}

\author{Bin Yan}
\affiliation{Theoretical Division, Los Alamos National Laboratory, Los Alamos, New Mexico 87545, USA}
\affiliation{Center for Nonlinear Studies, Los Alamos National Laboratory, Los Alamos, New Mexico 87545, USA}
\author{Nikolai A. Sinitsyn}
\affiliation{Theoretical Division, Los Alamos National Laboratory, Los Alamos, New Mexico 87545, USA}

\begin{abstract}
    Channel-state duality is a central result in quantum information science. It refers to the correspondence between a dynamical process (quantum channel) and a static quantum state in an enlarged Hilbert space. Since the corresponding dual state is generally mixed, it is described by a Hermitian matrix. In this article, we present a randomized channel-state duality. In other words, a quantum channel is represented by a collection of $N$ pure quantum states that are produced from a random source. The accuracy of this randomized duality relation is given by $1/N$, with regard to an appropriate distance measure. For large systems, $N$ is much smaller than the dimension of the exact dual matrix of the quantum channel. This provides a highly accurate low-rank approximation of any quantum channel, and, as a consequence of the duality relation, an efficient data compression scheme for mixed quantum states. We demonstrate these two immediate applications of the randomized channel-state duality with a chaotic $1$-dimensional spin system.
\end{abstract}

\title{Randomized channel-state duality}
\maketitle

Quantum channels are the most general framework for describing dynamical quantum processes, from the time evolution of closed or open quantum systems to quantum communications between distant parties, and error corrections on quantum computers. One of the most powerful methods for investigating quantum channels is the so-called channel-state duality \cite{Jiang2013Channel,Bengtsson2006Geometry,Leifer2006Quantum,Skowronek2009Cones,Zyczkowski2004On}: For every quantum channel, there exists a quantum state corresponding to it. As a result, the dynamical information of the former can be fully encoded into the kinematic information of the latter \cite{Arrighi2004On}. Hitherto, channel-state duality has become a classic textbook result in quantum information science. It not only offers an elegant mathematical characterization of the structure of quantum channels \cite{Horodecki2003Entanglement,Korbicz2012,Korbicz2012}, but also has a profusion of implications and applications in various research areas, e.g., quantum process tomography \cite{Altepeter2003Ancilla,DAriano2003Imprinting}, non-local quantum correlations \cite{Acin2010Unified,Barnum2010Local}, or non-Markovian quantum dynamics \cite{Luo2012Quantifying}.

The dual state of a quantum channel ``lives'' in an enlarged bipartite Hilbert space. In other words, for a channel that accepts an input state from a Hilbert space of dimension $d_a$, and outputs a state with dimension $d_b$, its corresponding dual state is a bipartite quantum state with a Hilbert space dimension $d=d_a\times d_b$. Additionally, the dual state is in general a mixed quantum state, and is therefore described by a density matrix---a Hermitian matrix of dimension $d\times d$ dubbed the Choi matrix. The rank of this matrix is also referenced as the rank of the corresponding channel. 

Although a quantum channel has a precise Choi matrix representation, efficiently finding its low-rank approximate \cite{Hayden2004Randomizing,Aubrun2009On,Lancien2017Approx} still remains a challenging problem. Such an approximation is highly desirable, because it can significantly reduce the complexity of describing and assessing the channel's properties. On the other hand, as a consequence of the channel-state duality, this problem is equivalent to finding a low-rank matrix approximation of the channel's Choi matrix. The latter problem is of importance on its own \cite{Eckart1936,Markovsky2008,Markovsky2012,Ezzell2022}, which has relevance in areas even outside physics, such as engineering and data sciences .

\begin{figure}[b!]
    \centering
    \includegraphics[]{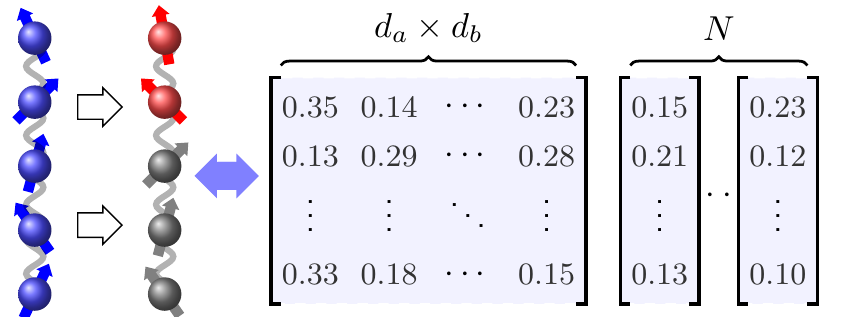}
    \caption{\textbf{Channel-state duality}. A quantum channel induced by the unitary evolution of an interacting spin chain system. The channel input is the state of the entire spin chain (blue), whose Hilbert space dimension is $d_a$. The output is the reduced state of a subsystem (red), with dimension $d_b$. Through channel-state duality, this channel can be represented by a (generally mixed) quantum state in a $d_a\times d_b$-dimensional Hilbert space, known as the Choi matrix. We show that the same channel can be described by a set of $N$ pure quantum states (hence vectors) of the same dimension, generated from random sources. Here $N$ determines the precision of the representation. } \label{fig:illustration}
\end{figure}

In this article, we introduce a \emph{randomized channel-state duality}. Instead of a single density matrix, we convert the channel to a set of $N$ pure states in the Hilbert space of the same dimension $d$ (Figure~\ref{fig:illustration}). These pure states are all produced from a random source of input. Here, the first moment of these pure states---averaged with respect to the probability distribution of the initial random input---creates an exact dual state (density matrix) of the quantum channel. Given that we employ $N$ random pure state realizations, the average of these pure states serves as a good approximation of the exact density matrix, with a precision (quantified by the variance of a proper distance measure) given by a factor of $1/N$.

\begin{figure*}[t!]
    \centering
    \includegraphics[]{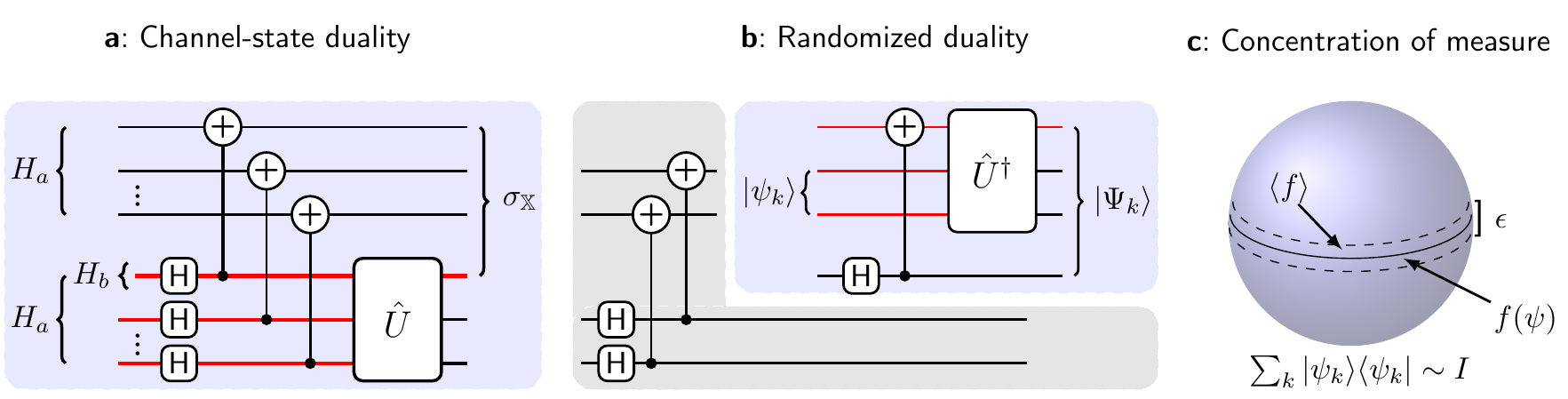}
    \caption{\textbf{Randomized channel-state duality}. (a) A unitary induced channel (red portion) can be represented by a quantum state $\sigma_\mathbb{X}$ in an extended Hilbert space, via the standard Jamio\l kowshi-Choi isomorphism (\ref{eq:Choi}). (b) Randomized channel-state duality maps the same channel to a set of quantum states $|\Psi_k\rangle$, $k=1,\cdots,N$, from a set of random input states $|\psi_k\rangle$, as defined in equation.~(\ref{eq:randomstate}). The mixture of a few random states $|\psi_k\rangle$ can approximate with very high accuracy the maximally mixed state. This is in analog to the concentration of measure (c) in high dimensional spaces, where typical values of a smooth function are close to the averaged value. Therefore, the random input $|\psi_k\rangle$ can be replaced with a system that is maximally entangled with an ancillary system (see the gray shaded region). The diagram in (b) together with the maximally entangled input state is equivalent to the Jamio\l kowshi-Choi representation, up to a local basis rotation $U^\dag\otimes U^\dag$ on the initial bipartite canonical maximally entangled state. }
   \label{fig:duality}
\end{figure*}

As a result, using $N$ $d$-dimensional vectors, we can approximate the precise dual state with a high degree of accuracy. $N$ is set to meet the desired precision. It is independent of, and much smaller than, the dimension $d$ for large systems.

\vspace{12pt}
\noindent\textbf{Randomized dual states}

Let us start by formulating the conventional channel-state duality. A quantum channel is formally defined as a linear map $\mathbb{X}: L(H^a)\rightarrow L(H^b)$ that transfers linear operators on Hilbert space $H^a$ to $H^b$, whose dimensions are respectively $d_a$ and $d_b$. $\mathbb{X}$ is demanded to be completely positive and trace-preserving. These properties guarantee the existence of the operator sum representation \cite{Kraus1970General} (Kraus representation) of channel $\mathbb{X}$, i.e.,
\begin{equation}
    \mathbb{X}\left(\rho\right) = \sum_{k=1}^r M_k \rho M_k^\dag,\quad \sum_k M_k^\dag M_k =I.
\end{equation}
Here, $I$ is the identity operator. The minimal value of $r$ is the Kraus rank (or Choi rank) of $\mathbb{X}$. 
To get the dual state of $\mathbb{X}$, consider a maximally entangled state $|\phi^+\rangle$ in the composed Hilbert space $H_a\otimes H_a$, and apply $\mathbb{X}$ to one of its subspace. This is also known as the Jamio\l kowshi-Choi isomorphism \cite{Pillis1967Linear,Jam1967Channel,Choi1975Completely}:
\begin{equation}\label{eq:Choi}
    \mathbb{X} \rightarrow \sigma_\mathbb{X} \equiv \mathbb{I}\otimes \mathbb{X}\left(|\phi^+\rangle\langle\phi^+|\right).
\end{equation}
Here, $\mathbb{I}$ is the identity map. The canonical maximally entangled state $|\phi^+\rangle \equiv \sum_i |ii\rangle/\sqrt{d_a} $ is represented in the computational basis. This correspondence is illustrated in Fig.~\ref{fig:duality}~a.  

The dual state $\sigma_\mathbb{X}$---known as the Choi matrix---is of dimension $d=d_a\times d_b$. It fully characterizes quantum channel $\mathbb{X}$, in the sense that any dynamical information of the channel can be extracted from the dual state alone. More precisely, for any Hermitian operators $A$ and $B$ that apply on $H_a$ and $H_b$, respectively, we have \cite{Arrighi2004On,Jiang2013Channel}
\begin{equation}\label{eq:ChoiDual}
    {\rm tr}\left[\mathbb{X}(A)B\right] = d_a\cdot {\rm tr}\left[\sigma_\mathbb{X}A^t \otimes B\right],
\end{equation}
where $A^t$ denotes the matrix transpose of $A$ in the computational basis. Note that the rank of the Choi matrix is identical to the Kraus rank of the corresponding channel. Therefore, a low-rank (approximate) representation of the Choi matrix directly gives rise to a low-rank representation of the channel, and vice versa.

\begin{figure*}[t!]
    \centering
    \includegraphics[]{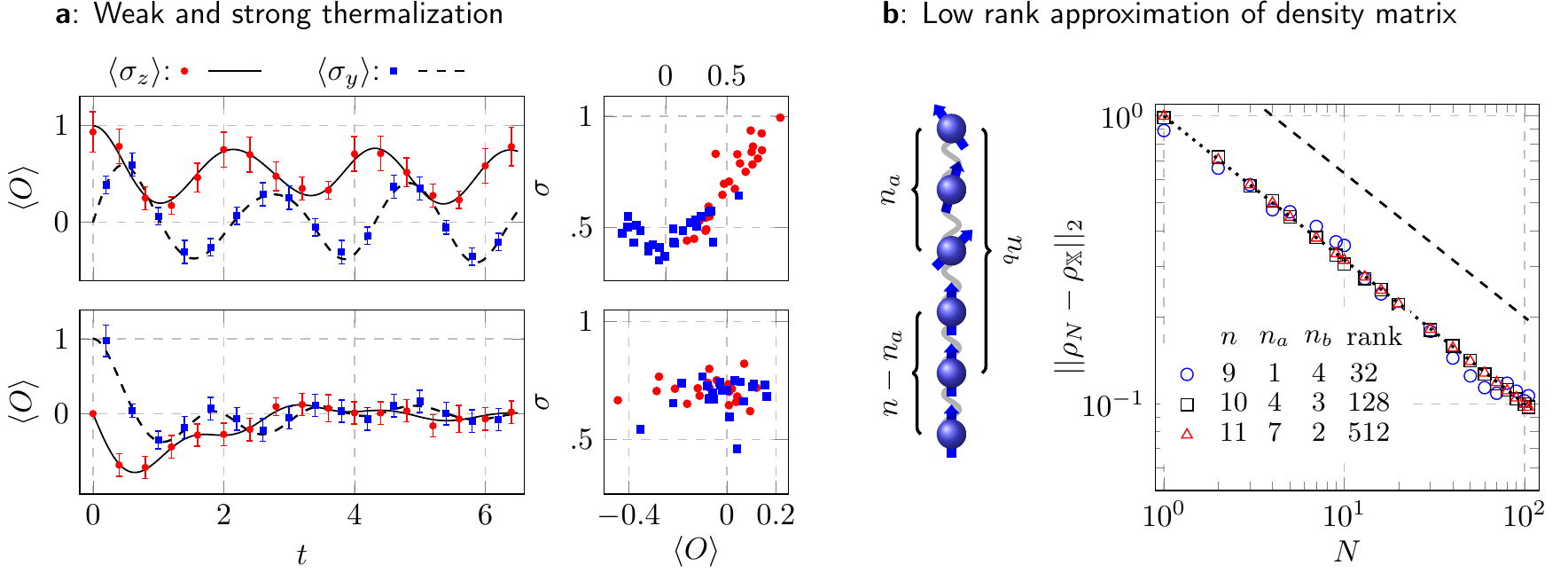}
    \caption{\label{fig:thermalization} (a) Weak and strong thermalization for a chaotic spin chain system with $12$ spins. Left: Time evolution of the expectation values of single spin observables. For weak (top) and strong (bottom) thermalization, the initial states of the spins are polarized in the $Z$ and $Y$ direction, respectively. Solid and dashed curves are direct numerical simulations of the evolution. Markers correspond to the averaged value evaluated with $N=200$ randomized dual states. Error bars show the confidence intervals of $3$-sigma [$3\sigma_N$ as defined in (\ref{eq:sigmaN})] of the data point. Right: Scatter plot of the standard deviation $\sigma$ defined in (\ref{eq:sigma}), which is below the predicted upper bound. (b) Trace distance between $\rho_{\mathbb{X}}$ in (\ref{eq:exactdual}) and its rank $N$ approximation $\rho^{\rm est}_{\mathbb{X}}$ (\ref{eq:dualest}), for various rank of $\rho_{\mathbb{X}}$ and $N$. The channel is generated by a unitary evolution of a $n$-site spin chain (\ref{eq:Ising}). The channel's input (output) space is the space of the first $n_a$ ($n_b$) spins. The dotted curve is the predicted averaged scaling $1/\sqrt{N}$. The dashed curve is away from the average value by the predicted upper bound of the standard deviation.}
\end{figure*}

For the sake of transparency, let us present the randomized channel-state duality for a special type of channel. We will generalize it later to generic channel. Consider a channel $\mathbb{X}$ induced by a unitary evolution $U$. The input Hilbert space $H_a$ matches the full dimension of the unitary, while the output Hilbert space $H_b$ is a subspace of $H_a$ (Figure~\ref{fig:illustration}). $\mathbb{X}$ can be formally define as
\begin{equation}
    \mathbb{X}(\rho) = {\rm tr}_{\bar{b}}\left[U\rho U^\dag\right],
\end{equation}
where ${\rm tr}_{\bar{b}}$ is the partial trace over the complement of $H_b$. We then map $\mathbb{X}$ to a pure state, i.e.,
\begin{equation}\label{eq:randomstate}
    \mathbb{X} \rightarrow |\Psi\rangle \equiv \mathbb{I}\otimes U^\dag \left(|\phi^+\rangle \otimes |\psi\rangle\right).
\end{equation}
This transformation is illustrated in Fig.~\ref{fig:duality}~b (blue shaded area).
Here, $|\phi^+\rangle$ is the canonical maximally entangled state in the bipartite Hilbert space $H_b\otimes H_b$---tensor product between the output Hilbert space and an ancillary Hilbert space of the same dimension. $|\psi\rangle$ is a random state on the complement of $H_b$. Note that the identity map applies on one subsystem of $|\phi^+\rangle$. $U^\dag$ applies to the other subsystem of $|\phi^+\rangle$ together with $|\psi\rangle$ Therefore, the resulting pure state has the same dimension $d$ as the Choi matrix. We also assume that the initial state $|\psi\rangle$ is drawn from an ensemble that forms a quantum state $2$-design \cite{Ambainis2007Quantum}. With respect to the probability distribution of the input ensemble, the first moment of the output state $|\Psi\rangle$ is a density matrix
\begin{equation}\label{eq:exactdual}
    \rho_\mathbb{X} \equiv \int d\psi~ |\Psi\rangle\langle\Psi|.
\end{equation}
Here, the integral is performed with respect to the probability measure of the initial random input state $|\psi\rangle$. This density matrix provides an exact characterization of channel $\mathbb{X}$, similar to the Choi matrix, through,
\begin{equation}\label{eq:characterization}
    {\rm tr}\left[\mathbb{X}(A)B\right] = d_a \cdot {\rm tr} \left[\rho_\mathbb{X}A \otimes B^t\right].
\end{equation}
Therefore, we get a new channel-state duality with the exact dual state $\rho_{\mathbb{X}}$.

Rigorous proof of the above duality relation is delegated to supplemental information. We now offer a more heuristic explanation: If one applies the standard Jamio\l kowshi-Choi isomorphism (\ref{eq:Choi}) not to $|\phi^+\rangle$, but to a maximally entangled state in a rotated basis other than the computational basis, i.e., $U^\dag\otimes U^\dag|\phi^+\rangle$, one gets a new transformation represented by a circuit diagram shown in Fig.~\ref{fig:duality}~b including the grey shaded area. For an observer who only has access to the space of the final output states $|\Psi\rangle$, the reduced state of one subsystem of the bipartite maximally entangled state is indistinguishable from the maximally mixed state. Hence, one can replace the maximally entangled input state (grey area in Fig.~\ref{fig:duality}~b) with the maximally mixed state, which can be further approximated by a collection of random pure states $|\psi_k\rangle$.

From this point of view, the exact density matrix $\rho_{\mathbb{X}}$ is not special compared to the standard Choi matrix $\sigma_{\mathbb{X}}$. In fact, as evidenced by their duality relations (\ref{eq:ChoiDual}) and (\ref{eq:characterization}), they are connected by a global transpose, which is an anti-unitary operation. However, the crucial point is that one can approximate $\rho_{\mathbb{X}}$ with $N$ realizations of the output pure state $|\Psi\rangle$, whose average serves as a good estimator of $\rho_{\mathbb{X}}$:
\begin{equation}\label{eq:dualest}
    \rho_\mathbb{X}^{\rm est} \equiv \frac{1}{N} \sum_{k=1}^N ~ |\Psi_k\rangle\langle\Psi_k|.
\end{equation}
As will be seen, we can achieve a high accurate approximation with only a relatively small number $N$.

\vspace{12pt}
\noindent\textbf{Bounding the variance}

The idea underlying the above low-rank approximation is the typicality of quantum states among a random ensemble. In our case, expectation values of observables evaluated on a single random dual state realization $|\Psi\rangle$ are highly likely to be around the averaged values of many realizations (Figure~\ref{fig:duality}~c). Quantitatively, the averaged distance between the exact dual state $\rho_{\mathbb{X}}$ and the estimator $\rho_{\mathbb{X}}^{\rm est}$ with $N$ pure states can be bounded as (supplemental information)
\begin{equation}
    \int d\bm{\psi}~||\rho_\mathbb{X}^{\rm est} -\rho_\mathbb{X}||_2 \le \sqrt{\frac{1}{N}}.
\end{equation}
Here $||X||_2\equiv\sqrt{{\rm tr}XX^\dag}$ is the Hilbert-Schmidt norm, and $ d\bm{\psi}\equiv \prod_kd\psi_k$
Moreover, the variance of the distance is suppressed by $N$ as well, i.e.,
\begin{equation}
    \int d\bm{\psi}~||\rho_\mathbb{X}^{\rm est} -\rho_\mathbb{X}||^2_2 \le \frac{1}{N}.
\end{equation}
This gives an upper bound $1/\sqrt{N}$ for the standard deviation of the distance. Since $\rho_{\mathbb{X}}^{\rm est}$, as the first moment of $N$ pure states, is a density matrix of rank at most $N$, we immediately get a low-rank approximation of the exact dual state $\rho_{\mathbb{X}}$

In many situations, it would be more convenient to directly work with observables rather than the dual states. To this end, we re-express the duality relation (\ref{eq:characterization}) as
 \begin{equation}\label{eq:operatordual}
    {\rm tr}\left[\mathbb{X}(A)B\right] = d_a \cdot \int d\psi~\langle\Psi|A \otimes B^t|\Psi\rangle.
\end{equation}
Here, on the right-hand side, the integrand multiplied by $d_a$ can be viewed as a random variable, whose average equals the measurement result of channel $\mathbb{X}$ on the left-hand side. Variance of this random variable is bounded by (see supplemental information)
\begin{equation}\label{eq:sigma}
    \sigma^2 \le C\cdot d_a^2 \cdot {\rm Var}\left[\mathbb{X}(A)B\right],
\end{equation}
where $C=1/(d_c+1)<1$.  ${\rm Var}\left[X\right]$ is the \emph{intrinsic variance} of operator $X$, defined with respect to the maximally mixed state, i.e.,
\begin{equation}
    {\rm Var}\left[X\right] \equiv d_a^{-1}{\rm tr}XX^\dag-d_a^{-2}{\rm tr}X{\rm tr}X^\dag.
\end{equation}
Note that the upper bound of $\sigma^2$ contains an extra factor $d_a^2$. This factor appears in the square of the mean as well, i.e.,
\begin{equation}
    \left|{\rm tr} \left[\mathbb{X}(A)B\right] \right|^2= d_a^2 \cdot \left|{\rm E}\left[\mathbb{X}(A)B\right]\right|^2,
\end{equation}
where ${\rm E}\left[X\right]$ is the expectation value of operator $X$, defined again with respect to the maximally mixed state, i.e.,
\begin{equation}
   {\rm E}\left[X\right] \equiv d_a^{-1} {\rm tr} X.
\end{equation}
Therefore, in our random approximation, the ratio between the variance and the square of the mean is fundamentally bounded by that of the operator $UAU^\dag B$.
Note also that $\sigma^2$ is the variance of the measurement result that corresponds to a single realization of $|\Psi\rangle$. The variance of the average of $N$ realizations of the dual states is further suppressed by a factor of $1/N$, i.e,
\begin{equation}\label{eq:sigmaN}
    \sigma^2_N = \sigma^2 /N. 
\end{equation}
The bound of variance $\sigma^2$ can be improved in some cases. For instance, as discussed in supplemental information, when $A$ is a rank-$1$ projector and $B$ is a positive operator, $\sigma^2$ is directly bounded by the square of the mean. This case is of particular interest since it corresponds to the scenario of pure initial input state and general POVM measurements.

To elaborate with a concrete example, let us consider a practical situation where the randomized channel-state duality can offer a boost of computational advantages. Assume that we are given an isolated system and that we would like to calculate a local observable's expectation values at a specific time, for various initial conditions. In this case, the channel is induced by fixed unitary evolution. One can generate a collection of $N$ pure dual states only once, with which the measurement of local observables can be computed for various initial conditions, without having to evaluate the unitary evolution every time.

The system we studied is an Ising spin chain with both transverse and longitudinal magnetic field. The Hamiltonian is
\begin{equation}\label{eq:Ising}
    H = - \sum_i \sigma_i^z\sigma_{i+1}^z - g \sum_i \sigma_i^x - h \sum_i \sigma_i^z,
\end{equation}
where $\sigma_i$ are the Pauli matrices. For $h$ and $g$ that are not vanishing simultaneously, this system is chaotic and hence exhibits thermalization. However, depending on the initial condition, the thermalization can be weak or strong \cite{Banuls2011Strong}. That is, the expectation values of local observables on average saturate to their thermal values, but with large and small fluctuations, respectively. We have simulated the expectation values of local Pauli operators for both weak and strong thermalization, and compared the result evaluated from the randomized dual states with that directly computed from exact unitary evolution. The fields are fixed at $g=1.05$ and $h=0.5$. Here, in equation (\ref{eq:operatordual}), operator $A$ is then the initial density matrix of the system, and $B$ is $\sigma_1^{z(y)}$. In this case, $\sigma$ in equation (\ref{eq:sigma}) is bounded by $\sigma\le\sqrt{2}$. Figure~\ref{fig:thermalization}~a demonstrates that the random dual states predict the same result as the exact unitary evolution, with deviations that follow exactly our prediction (\ref{eq:sigma}) and (\ref{eq:sigmaN}). 

\vspace{12pt}
\noindent\textbf{Generalizations}

To generalize the randomized channel-state duality to generic quantum channels, $\mathbb{X}: L(H_a) \rightarrow L(H_b)$, note that $\mathbb{X}$ can be dilated to a unitary channel via the Stinespring dilation theorem \cite{Stinespring1955Positive}. Namely, the input Hilbert space is enlarged by an ancillary system, whose initial state is fixed to a pure state, denoted as $|0\rangle$. One can then generate the randomized dual states for this enlarged unitary channel. The duality relation (\ref{eq:operatordual}) becomes
\begin{equation}
    {\rm tr} \left[\mathbb{X}(A)B\right] = d_U\cdot \int d\bm{\psi} \langle\Psi| A\otimes|0\rangle\langle0|\otimes B^t |\Psi\rangle,
\end{equation}
where $d_U$ is the dimension of the dilated unitary. Let us decompose the dual state $|\Psi\rangle$ as
\begin{equation}
    |\Psi\rangle = c_0 |0\rangle\otimes|\Phi\rangle + |\Tilde{\Psi}\rangle,
\end{equation}
where $|\Phi\rangle$ lives in the Hilbert space that support $A\otimes B$, and $|\Tilde{\Psi}\rangle$ is orthogonal to $|0\rangle$. $c_0^2 = d_a/d_U$ is a normalization factor. With this, we recover the duality relation
\begin{equation}
    {\rm tr} \left[\mathbb{X}(A)B\right] = d_a\cdot  \int d\bm{\psi} \langle\Phi| A\otimes B^t |\Phi\rangle.
\end{equation}
Note that $|\Phi\rangle$ has the same dimension as the the Choi matrix of $\mathbb{X}$. Their average forms an exact dual state $\rho_{\mathbb{X}}=\int d\bm{\psi} |\Phi\rangle\langle\Phi|$, which can be approximated with $N$ realizations of $|\Phi\rangle$. The choice of $c_0$ guarantees that $\rho_{\mathbb{X}}$ is normalized. To test this, consider the Ising spin system again. This time, the channel input and output are both chosen as subsystems of the entire spin chain (in this case, the channel dilation is already known, which is the unitary evolution of the entire system). Figure~\ref{fig:thermalization}~b shows the distance between the exact dual matrix and its approximates with various $N$. The scaling of the distance as well as the deviations follow our prediction.

Another benefit of using the randomized dual (pure) states is to consider their higher-order moments. In contrast to the first moment, which is approximate to the exact dual state of the channel, the higher-order moments contain information beyond the standard channel-state duality. This can be used to extract higher-order correlations of the quantum channel. For example, for the unitary induced channel studied in the previous section, and when operator $B=\Pi_B$ is a rank-$1$ projector in the computational basis, the second-order average of the observables is
\begin{equation}\label{eq:otoc}
\begin{aligned}
            &d^2_a\cdot \int d\psi d\psi'~ \left|\langle\Psi| A\otimes \Pi_B |\Psi'\rangle\right|^2 =
         {\rm tr} \left[(UAU^\dag \Pi_B)^2 \right],
\end{aligned}
\end{equation}
where the right-hand side is the out-of-time order correlation function \cite{Kitaev2015,Larkin1969}. This quantity has been used to study information scrambling \cite{Swingle2018Unscrambling} and diagnose chaos \cite{Maldacena2016Bound} in quantum dynamics. Equation (\ref{eq:otoc}) provides a practical strategy for measuring the OTOC without driving the system through forward and backward evolution loops, similar to the approach in \cite{Vermersch2019Probing}. Implications of this observation, as well as general higher-order channel-state dualities, deserve further investigation.

Let us also remark on the randomness and typicality of quantum state ensembles, which is the primary mechanism behind the randomized channel-state duality. This phenomenon is not new to physicists. For example, it has been used to replace the equal a \emph{priori} probability postulate in statistical physics \cite{Popescu2006Entanglement,Goldstein2006Canonical}, derive universal behaviors of quantum chaotic systems \cite{Roberts2017-tt,Yan2020Information,Yan2020Recovery}, establish fundamental limitations of quantum machine learning \cite{McClean2018Barren,Holmes2021Barren}, or probe entanglement with randomized measurement \cite{Brydges2019Probing}. Here, our findings suggest that one can employ randomness of quantum states passively as a resource to encode and process quantum information as well. 

\vspace{12pt}
\noindent\textbf{Acknowledgement}

This work was supported in part by the U.S. Department of Energy, Office of Science, Office of Advanced Scientific Computing Research, through the Quantum Internet to Accelerate Scientific Discovery Program, and in part by U.S. Department of Energy under the LDRD program at Los Alamos. B.Y. also acknowledges support from the Center for Nonlinear Studies.

\vspace{12pt}
\noindent\textbf{References}

\bibliography{references}

\clearpage
\onecolumngrid

\setcounter{page}{1}
\setcounter{section}{0}
\setcounter{equation}{0}
\setcounter{figure}{0} 

\vspace{30pt}
\begin{center}
{\large Supplemental information for ``Randomized channel-state duality''}\\
\vspace{8pt}
Bin Yan$^{1,2}$ and Nikolai A. Sinitsyn$^1$\\
\vspace{3pt}
{\small $^1$ Theoretical Division, Los Alamos National Laboratory, New Mexico 87545, USA}\\
{\small $^1$ Center for Nonlinear Studies, Los Alamos National Laboratory, New Mexico 87545, USA}
\end{center}

\tableofcontents

\section{Establishing the duality --- first order correlation}

\begin{figure*}[h!]
    \centering
    \includegraphics[width=\textwidth]{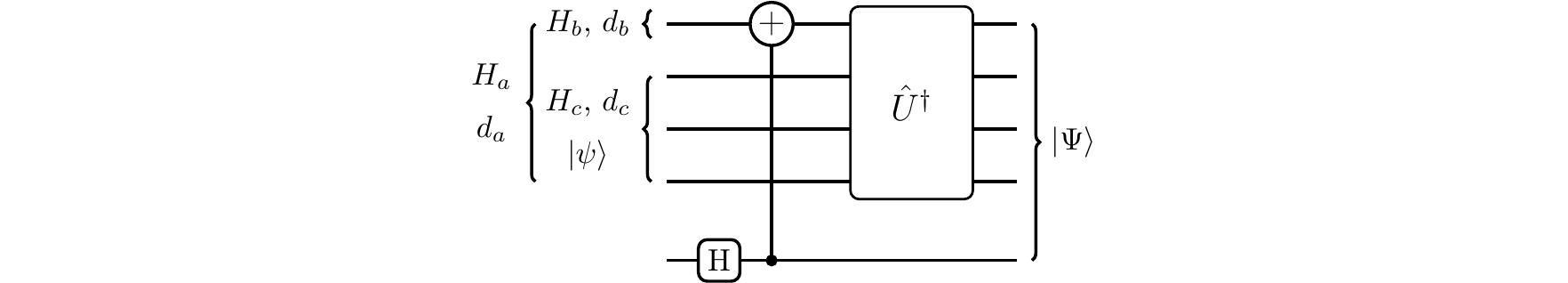}
    \caption{Randomized channel-state duality for channel $\mathbb{X}: L(H_a)\rightarrow L(H_b)$ induced by unitary $U$.}
    \label{fig:si_duality}
\end{figure*}

In this section, we prove that, for a quantum channel $\mathbb{X}$ directly induced by unitary $U$, the first moment of the random dual states introduced in the main text is an exact dual state of the channel. 

Let us first fix the convention. As illustrated in Fig.~\ref{fig:si_duality}, the channel maps quantum states from Hilbert space $H_a$ of dimension $d_a$, to Hilbert space $H_b$ of dimension $d_b$. This is a unitary induced channel in the sense that the input dimension $d_a$ is the dimension of unitary, and $H_b$ is a subsystem of $H_a$. The channel is formally defined as
\begin{equation}
    \mathbb{X}(\rho) = {\rm tr}_{\bar{b}}\left[U\rho U^\dag\right],
\end{equation}
where ${\rm tr}_{\bar{b}}$ is the partial trace over the complement of $H_b$, i.e., $H_c$. Equivalently, for Hermitian operators $A$ and $B$ that apply on $H_a$ and $H_b$, respectively,
\begin{equation}
   {\rm tr} \left[\mathbb{X}(A)B\right] = {\rm tr}\left[U\rho U^\dag B\right],
\end{equation}

For this channel, the random dual states are generated as
\begin{equation}
    |\Psi\rangle = \mathbb{I} \otimes U^\dag \left(|\phi^+\rangle\otimes|\psi\rangle\right),
\end{equation}
where $\mathbb{I}$ is the identity map. $|\psi\rangle$ is a random state drawn from an ensemble that is a quantum state $2$-design. $|\psi\rangle$ is a state on the complement of $H_b$, which is denoted as $H_c$ with dimension $d_c$. $|\phi^+\rangle \equiv \sum_i |ii\rangle/ \sqrt{d_b}$ is the maximally entangled state in the computational basis that lives in $H_b\otimes H_b$---a tensor product space between the subsystem $H_b$ and an ancillary copy of $H_b$.

The first moment of the random dual state $|\Psi\rangle$ reads
\begin{equation}
    \rho_\mathbb{X} \equiv \int d\psi |\Psi\rangle\langle\Psi|,
\end{equation}
where the integral is performed with respect to the measure of $|\psi\rangle$ in the initial random ensemble (in our case, a $2$-design). Our aim is to establish an exact duality relation introduced in the main text, i.e.,
\begin{equation}
    {\rm tr}\left[\mathbb{X}(A)B\right] = d_a \cdot {\rm tr} \left[\rho_\mathbb{X}A \otimes B^t\right].
\end{equation}
On the right hand side of the above equation, $\rho_\mathbb{X}$ lives on the Hilbert space $H_a\otimes H_b$. Operator $A$ applies on $H_a$, and $B$ applies on the ancillary copy of $H_b$.
This condition is equivalent to
\begin{equation}
    {\rm tr}\left[UAU^\dag B^t\right] = d_a \cdot \int d\psi \langle\Psi|A \otimes B|\Psi\rangle.
\end{equation}
The right hand side of the above equation (besides the pre-factor $d_a$) is the average of the measurement value of observable $A\otimes B$ on $|\Psi\rangle$, denoted as
\begin{equation}
\begin{aligned}
     \mu_1 \equiv &\int d\psi~ \langle\Psi| A\otimes B |\Psi\rangle\\
    =& \int d\psi~ {\rm tr} \left( |\psi\rangle \langle \psi|\otimes|\phi^+\rangle\langle\phi^+|  U A U^\dag \otimes B  \right).   
\end{aligned}
\end{equation}
To evaluate this quantity further, note that the maximally entangled state $|\phi^+\rangle$ satisfies the invariance property
\begin{equation}
    X\otimes Y |\phi^+\rangle = XY^t\otimes I |\phi^+\rangle = I \otimes YX^t |\phi^+\rangle,
\end{equation}
where $I$ is the identity operator. This allows us to move operator $B$, which acts on the ancillary copy of $H_b$ system, to the subsystem $H_b$, i.e.,
\begin{equation}
    I\otimes B |\phi^+\rangle = B^t\otimes I |\phi^+\rangle.
\end{equation}
Hence, the average simplifies to
\begin{equation}
\mu_1 = \int d\psi~ {\rm tr} \left( |\psi\rangle \langle \psi| \otimes |\phi^+\rangle \langle \phi^+|  U A U^\dag B^t \otimes I  \right).
\end{equation}
Since $UAU^\dag B^t$ only applies to $|\psi\rangle$ and one subsystem $H_b$ of the bipartite entangled state $|\phi^+\rangle \in H_b\otimes H_b$ , only this subsystem is relevant in the above trace evaluation. We can then replace $|\phi^+\rangle$ with its reduced density of state on the subsystem $H_b$ that supports $UAU^\dag B^t$, which is a maximally mixed state $I/d_b$. Therefore, $\mu_1$ further simplifies to
\begin{equation}
\mu_1 = \frac{1}{d_b} \int d\psi~ {\rm tr} \left( |\psi\rangle \langle \psi|  U A U^\dag B^t \right).
\end{equation}

We consider that the initial random input state $|\psi\rangle$ is generated from a fixed reference state $|0\rangle$ by random unitary $V$ that is drawn from a unitary $2$-design ensemble, i.e., $|\psi\rangle=V|0\rangle$. $|\psi\rangle$ generated in this manner automatically form a quantum state $2$-design. With this, we can replace the integral over $|\psi\rangle$ with an integral over $V$ with respect to its measure on the $2$-design ensemble, i.e.,
\begin{equation}
\begin{aligned}
           \mu_1 =& \frac{1}{d_b} {\rm tr} \int dV~ V|0\rangle \langle 0|V^\dag U A U^\dag B^\dag\\
            =& \frac{1}{d_a} {\rm tr} \left( U A U^\dag B^t\right),
\end{aligned}
\end{equation}
where we used the first-order Haar average
\begin{equation}
    \int dV~V|0\rangle\langle 0|V^\dag = I/d_c.
\end{equation}
Since $V$ is drawn from a unitary $2$-design, its first moment average equal the Haar average. Note also that $d_c$ is the dimension of $V$, and $d_a=d_bd_c$. Hence, we obtain the desired channel-state duality
\begin{equation}
    {\rm tr} \left( U A U^\dag B^t\right) = d_a\mu_1 =  d_a \cdot \int d\psi \langle\Psi|A \otimes B|\Psi\rangle = {\rm tr} \left[\rho_\mathbb{X}A \otimes B\right].
\end{equation}

\section{Bounding the variance}
In the previous section, we have established the exact channel-state duality for the first moment of the random dual states,
\begin{equation*}
    \rho_\mathbb{X} = \int d\psi |\Psi\rangle\langle\Psi|.
\end{equation*}
With $N$ realizations of the pure random dual states $|\Psi\rangle$, we achieve an estimate of the exact dual state, that is
\begin{equation*}
    \rho_\mathbb{X}^{\rm est} = \frac{1}{N}\sum_{k=1}^N |\Psi_k\rangle\langle\Psi_k| \approx \rho_\mathbb{X}.
\end{equation*}
Here, we use two methods to quantify the accuracy of this approximation. One is via direct computation of the distance between the exact and the approximate states, the other is to quantify the variance of the observable expectation value predicted by $\rho_\mathbb{X}^{\rm est}$. We first present the second approach.
 
\subsection{Variance for observables}

\subsubsection{General upper bound}
In this section, we derive a tight upper bound of the variance for observables. For Hermitian operators $A$ and $B$ that apply on $H_a$ and $H_b$, respectively, the duality relation reads 
\begin{equation}
    {\rm tr}\left[\mathbb{X}(A) B^t\right] = d_a \cdot \int d\psi \langle\Psi|A \otimes B|\Psi\rangle.
\end{equation}
Here, $\langle\Psi|A \otimes B|\Psi\rangle$ is the expectation value of $A\otimes B$ for a single random dual state $|\Psi\rangle$. It can be viewed as a random variable, whose average value equals the quantum channel prediction on the left-hand side of the above equation (upto a pre-factor $d_a$). We would like to quantify the variance of this random variable, i.e.,
\begin{equation}
\begin{aligned}
            \sigma^2 \equiv & \int d\Psi~ \left|\langle\Psi| A\otimes B |\Psi\rangle\right|^2 - \left(\int d\Psi~ \langle\Psi| A\otimes B |\Psi\rangle\right)^2\\
=&\mu_2 - \mu_1^2.
\end{aligned}
\end{equation}
Here, $\mu_2$ is defined as
\begin{equation}
\begin{aligned}
       \mu_2 \equiv& \int d\psi~ \left|\langle\Psi| A\otimes B |\Psi\rangle\right|^2\\
       =& \int d\psi~ \left|{\rm tr} \left( |\psi\rangle \langle \psi|\otimes|\phi^+\rangle\langle\phi^+|  U A U^\dag \otimes B  \right)\right|^2. 
\end{aligned}
\end{equation}
We would like to stress that, in the main text, we considered the random variable $d_a\cdot\langle\Psi|A \otimes B|\Psi\rangle$, which differs from the convention here by a factor of $d_a$. Therefore, the variance defined in the main text is the one considered here multiplied by $d^2_a$.

Let us evaluate this term first. With the same trick we used in the previous section, the bi-partite entangled state $|\phi^+\rangle \in H_b\otimes H_b$ can be replaced by a maximally mixed state $I/d_b$ on $H_b$, i.e., 
\begin{equation}
       \mu_2 = \frac{1}{d^2_b} \int d\psi~ \left[ {\rm tr} \left( |\psi\rangle \langle \psi|  UAU^\dag B^t \right)\right]^2.
\end{equation}
Let us write $\mu_2$ in a more convenient form
\begin{equation}
       \mu_2 = \frac{1}{d^2_b} \int d\psi~  {\rm tr} \left( |\psi\rangle \langle \psi|  UAU^\dag B^t \right) {\rm tr} \left[ |\psi\rangle \langle \psi|  (UAU^\dag B^t )^\dag \right],
\end{equation}
and decompose $ UAU^\dag B^t$ as
\begin{equation}
   UAU^\dag B^t  = \sum_k P_k\otimes Q_kB^t,
\end{equation}
where $P_k$ and $Q_k$ are Hermitian operators on $H_c$ and $H_b$, respectively. $\{P-k\}$ form an orthonormal frame, i.e.,
\begin{equation}
    {\rm tr} P_kP_{k'}^\dag = \delta_{k,k'}.
\end{equation}
To see this, choose two orthornomal Hermitian frames $P_k$ and $\Tilde{Q}_k'$, and decompose $UAU^\dag$ as
\begin{equation}
   UAU^\dag  = \sum_{k,k'} c_{k,k'}P_k\otimes \tilde{Q}_{k'} = \sum_k P_k\otimes \sum_{k'}c_{k.k'}\tilde{Q}_{k'}.
\end{equation}
Since $A$ is Hermitian, $c_{k,k'}$ are real numbers. Therefore, $Q_k\equiv \sum_{k'}c_{k.k'}\tilde{Q}_{k'}$ are Hermitian (but not orthonormal).

With this decomposition, we can evaluate $\mu_2$ as
\begin{equation}
\begin{aligned}
        \mu_2 =& \frac{1}{d^2_b} \int d\psi~ \sum_{k,k'} {\rm tr} \left( |\psi\rangle \langle \psi| P_k\otimes Q_kB^t \right) {\rm tr} \left( |\psi\rangle \langle \psi| P^\dag_{k'}\otimes B^t Q^\dag_{k'} \right)\\
        =& \frac{1}{d^2_b} \int d\psi~ \sum_{k,k'} {\rm tr} \left( |\psi\rangle \langle \psi| P_k  \right) {\rm tr} \left( |\psi\rangle \langle \psi| P^\dag_{k'}\right) {\rm tr} Q_k B^t {\rm tr} B^t Q^\dag_{k'}\\
        =& \frac{1}{d^2_b} \int d\psi~ \sum_{k,k'} {\rm tr} \left( |\psi\rangle \langle \psi| P_k |\psi\rangle \langle \psi| P^\dag_{k'}\right) {\rm tr} Q_k B^t {\rm tr} B^t Q^\dag_{k'}\\
\end{aligned}
\end{equation}

Again, via $|\psi\rangle = V|0\rangle$, we replace the integral over $|\psi\rangle$ with an integral over a unitary $2$-design, which can be computed with the aid of the Weingarten
function for Haar random unitaries. We put the resulting formula here:
\begin{equation}
\int dV\ \ V^\dag X V Y V^\dag Z V
= \left[ \frac{{\rm tr}X\ {\rm tr}Z}{D^2-1} - \frac{{\rm tr}(XZ)}{D(D^2-1)}\right] Y 
+ \left[\frac{{\rm tr}(XZ)\ {\rm tr}Y}{D^2-1} - \frac{{\rm tr}X\ {\rm tr}Z\ {\rm tr}Y}{D(D^2-1)} \right] I,
\end{equation}
where $X, Y$ and $Z$ are arbitrary operators. $D$ is the dimension of the unitary $V$. In our case, since $V$ is drawn from a unitary $2$-design, its second moment average equals the above Haar average. Therefore,
\begin{equation}\label{eq:si_mu2}
\begin{aligned}
        \mu_2  =&\frac{1}{d^2_b} \sum_{kk'} \left({\rm tr} \int dV~ V|0\rangle \langle 0|V^\dag P_k V|0\rangle \langle 0|V^\dag P^\dag_{k'} \right) {\rm tr}Q_k B^t{\rm tr} B^t Q^\dag_{k'}\\
        =&\frac{1}{d^2_b}\frac{1}{d_c(d_c+1)} \sum_{kk'}  \left( {\rm tr}  P_kP^\dag_{k'} + {\rm tr} P_k{\rm tr} P^\dag_{k'}\right){\rm tr}Q_k B^t {\rm tr}B^t Q^\dag_{k'}.
\end{aligned}
\end{equation}
The second term in the parentheses of the above equation results in a quantity that is proportional to the square of $\mu_1$, i.e.,
\begin{equation}
    \frac{1}{d_b^2d_c(d_c+1)}\sum_{kk'}{\rm tr} P_k{\rm tr} P^\dag_{k'}{\rm tr}Q_k B^t {\rm tr}B^t Q^\dag_{k'} = \frac{d_c}{d_c+1} \frac{1}{d_a^2} \left[ {\rm tr}\left(UAU^\dag B^t\right)\right]^2 = \frac{d_c}{d_c+1}\mu_1^2,
\end{equation}
where we have used $d_a=d_bd_c$. The first term in the parentheses can be bounded as
\begin{equation}
\begin{aligned}
    &\frac{1}{d_b^2}\frac{1}{d_c(d_c+1)}\sum_{kk'}{\rm tr} P_k P^\dag_{k'}{\rm tr}Q_k B^t {\rm tr}B^t Q^\dag_{k'}\\
    \le & \frac{1}{d_b^2}\frac{d_b}{d_c(d_c+1)}\sum_{k}{\rm tr} P_k P_{k}^\dag{\rm tr}\left[Q_k B^t (Q_{k}B^t)^\dag\right] \\
      =& \frac{1}{d_a}\frac{1}{d_c+1} {\rm tr}\left[\sum_k P_k\otimes Q_k B^t \sum_{k'} P^\dag_{k'}\otimes (Q^\dag_{k'}B^t)^\dag\right]\\
      =&\frac{1}{d_a}\frac{1}{d_c+1} {\rm tr} \left[UAU^\dag B^t\left(UAU^\dag B^t\right)^\dag\right].
\end{aligned}
\end{equation}
In the second line of above equation, we have used the orthogonality of $\{P_k\}$, and the inequality is due to combining the two trace terms involving $B^t$.  Namely, for operator $X$,
\begin{equation}
    {\rm tr} X {\rm tr} X^\dag \le D\cdot {\rm tr} XX^\dag,
\end{equation}
where $D$ is the dimension of $X$. To establish this, note that the left-hand side is bounded in terms of the trace norm (nuclear norm) of $X$, i.e.,
\begin{equation}
    {\rm tr}X, {\rm tr}X^\dag \le {\rm tr}\sqrt{X^\dag X}\equiv ||X||_*,
\end{equation}
and the right-hand side is the square of the Hilbert-Schmidt norm (Frobenius norm), i.e.,
\begin{equation}
    {\rm tr} XX^\dag \equiv ||X||_2^2.
\end{equation}
These two norms satisfy
\begin{equation}
    ||X||_* \le \sqrt{D}\cdot ||X||_2.
\end{equation}
Therefore, the variance becomes
\begin{equation}
\begin{aligned}
       \sigma^2 \le  \frac{1}{d_c+1}\left\{\frac{1}{d_a}{\rm tr} \left[UAU^\dag B^t\left(UAU^\dag B^t\right)^\dag\right] - \frac{1}{d^2_a}\left[ {\rm tr}\left(UAU^\dag B^t\right)\right]^2\right\}.
\end{aligned}
\end{equation}
Here, the term in the curly brackets can be interpreted as the intrinsic variance of the operator $UAU^\dag B^t$ (evaluated with respect to a maximally mixed state $I/d_a$), defined as
\begin{equation}
    {\rm Var}\left[UAU^\dag B^t\right] \equiv {\rm tr} \left[UAU^\dag B^t\left(UAU^\dag B^t\right)^\dag I/d_a\right] - \left[ {\rm tr}\left(UAU^\dag B^t I/d_a\right)\right]^2.
\end{equation}
Therefore, 
\begin{equation}
    \sigma^2 \le C\cdot {\rm Var}\left[UAU^\dag B^t\right],
\end{equation}
where $C=1/(d_c+1)<1$. This is a general tight upper bound, in the sense that it can be saturated. For instance, when $A$ and $B$ are both identity operators, $\sigma^2 = 0$. 

Note that the square of the mean, $\mu_1^2$, is interpreted as the expectation value of the same operator $UAU^\dag B^t$, with respect to the maximally mixed state $I/d_a$, i.e.,
\begin{equation}
    \mu_1^2 = \left[ {\rm tr}\left(UAU^\dag B^t I/d_a\right)\right]^2.
\end{equation}
The ``noise-to-signal'' ratio, $\sigma^2/\mu_1^2$, is therefore determined by that of the operator $UAU^\dag B^t$.

Since our protocol involves $N$ independent random states, we are interested in the ``$N$-averaged'' value:
\begin{equation}
    \frac{1}{N}\sum_{k=1}^N \langle\Psi_k| A\otimes B|\Psi_k\rangle = {\rm tr}\left[\rho_{\mathbb{X}}^{\rm est}A\otimes B \right].
\end{equation}
Here, $|\Psi_k\rangle$ are independent realizations of the random dual states. The overall variance $\sigma^2_N$ of the above quantity is therefore suppressed by a factor of $N$, i.e.,
\begin{equation}\label{sm:eq:sigmaN}
    \sigma^2_N = \frac{1}{N} \sigma^2 \le \frac{C}{N} {\rm Var}\left[UAU^\dag B^t\right].
\end{equation}

As an example, consider a system with $n$ qubits. In this case, $d_a=2^n$. As studied in the main text, $A$ is a rank-$1$ projector corresponding to the initial pure state of the system. $B$ is a single qubit Pauli operator. In this case, $d_c = 2^{n-1}$. The bound of $\sigma^2$ reduces to
\begin{equation}
\begin{aligned}
       \sigma^2 \le  \frac{1}{d_c+1}\frac{1}{d_a}{\rm tr} \left[UAU^\dag B^t\left(UAU^\dag B^t\right)^\dag\right] = \frac{1}{(d_c+1)d_a}\le \frac{2}{d_a^2}.
\end{aligned}
\end{equation}

Again, in the main text, the considered random variable is $d_a\cdot \langle \Psi|A\otimes B |\Psi\rangle$, rather than $ \langle \Psi|A\otimes B |\Psi\rangle$ studied here. Therefore, the above bound of $\sigma^2$ translates to $\sigma^2\le 2$ in the main text.

\subsubsection{Rank-1 projectors}
The upper bound of variance $\sigma^2$ can be improved for specific cases. In this section, we derive a tighter bound for the case when the operators $A$ and $B$ are rank-$1$ projectors ($B$ can be further relaxed to any positive operator).

Let us start with the expression of $\mu_2$ we have obtained in equation~(\ref{eq:si_mu2}),
\begin{equation}
\begin{aligned}
        \mu_2 =&\frac{1}{d^2_b}\frac{1}{d_c(d_c+1)} \sum_{kk'}  \left( {\rm tr}  P_kP^\dag_{k'} + {\rm tr} P_k{\rm tr} P^\dag_{k'}\right){\rm tr}Q_k B^t {\rm tr}B^t Q^\dag_{k'},
\end{aligned}
\end{equation}
where $P_k$ and $Q_k$ are operators from the decomposition
\begin{equation}
    UAU^\dag = \sum_k P_k\otimes Q_k.
\end{equation}
Suppose $UAU^\dag = |s\rangle\langle s|$ is a rank-$1$ operator, we can decompose $|a\rangle$ in $H_c\otimes H_b$ as
\begin{equation}
    |s\rangle = \sum_i |i_c\rangle\otimes|i_b\rangle,
\end{equation}
where $|i_b\rangle$ is chosen as the the eigenbasis of $B$. $|i_c\rangle$ are states that are neither normalized nor orthogonal. We can then write $UAU^\dag$ as
\begin{equation}
\begin{aligned}
    UAU^\dag = \sum_{i,j} |i_c\rangle\langle j_c|\otimes|i_b\rangle\langle j_b|.
\end{aligned}
\end{equation}
With this, one can identify that
\begin{equation}
    P_k = |i_c\rangle\langle j_c|, \quad Q_k=|i_b\rangle\langle j_b|.
\end{equation}
Here, $k$ is a label for the pair $(i,j)$. Since $i_b$ is the eigenbasis of $B$, the only terms that survive in $\mu_2$ is those terms that have $k = (i,i)$. For those $k$, we have
\begin{equation}
    {\rm tr}  P_kP^\dag_{k'} \le  {\rm tr}  P_k {\rm tr}P^\dag_{k'}, \quad {\rm tr}Q_kB^t, {\rm tr}B^tQ_{k'}>0,
\end{equation}
where latter is because $B$ is a positive operator as assumed. Therefore,
\begin{equation}
\begin{aligned}
        \mu_2 \le& \frac{1}{d^2_b}\frac{2}{d_c(d_c+1)} \sum_{kk'}  {\rm tr} P_k{\rm tr} P^\dag_{k'} {\rm tr}Q_k B^t {\rm tr}B^t Q^\dag_{k'}\\
        =& \frac{2}{d_a^2} \left[ {\rm tr}\left(UAU^\dag B^t\right)\right]^2 = 2\mu_1^2.
\end{aligned}
\end{equation}
Hence, the variance is bounded by the square of the mean, i.e.,
\begin{equation}
    \sigma^2 \le \mu_1^2.
\end{equation}
This is a strong result: no matter how small the expectation value $|\mu_1|$ is, with $N$ realizations of the random dual states, we can always achieve a precision $\sigma \le |\mu_1|/\sqrt{N}$.

\subsection{Variance for the dual state}

We now switch to directly bounding the distance (Hilbert-Schmidt) between the exact dual state $\rho_\mathbb{X}$ and its estimate with $N$ pure random dual states $\rho_\mathbb{X}^{\rm est}$. It is convenient to evaluate the exact dual state further:
\begin{equation}
\begin{aligned}
    \rho_\mathbb{X} \equiv&  \int d\psi |\Psi\rangle\langle\Psi|\\
    =& \int d\psi~ U^\dag |\phi^+\rangle\langle\phi^+|\otimes |\psi\rangle\langle \psi|  U\\
    =&  \int dV~ U^\dag |\phi^+\rangle\langle\phi^+|\otimes V|0\rangle\langle 0|V^\dag  U\\
    =& \frac{1}{d_c} U^\dag \left(|\phi^+\rangle\langle\phi^+|\otimes I\right)  U,
\end{aligned}
\end{equation}
where in the last line we used again the first moment integral over Haar random unitaries. Note that, as illustrated in Fig.~\ref{fig:si_duality} in the first section, $|\phi^+\rangle \in H_b\otimes H_b$, $I$ is an identity operator on $H_c$, and the unitary $U$ applies on $H_a = H_b\otimes H_c$. With this expression, it is clear that 
\begin{equation}
    \rho_\mathbb{X}^2 = \frac{1}{d_c}\rho_\mathbb{X}.
\end{equation}
Another useful quantity is the average of the square of the difference between $\rho_\mathbb{X}$ and $\rho_\mathbb{X}^{\rm est}$, i.e., 
\begin{equation}
    \int d\bm{\psi}~ \left(\rho_\mathbb{X}^{\rm est}-\rho_\mathbb{X}\right)^2
    \equiv \int d\psi_1\cdots d\psi_N~\left(\frac{1}{N}\sum_{i=1}^N |\Psi_i\rangle\langle\Psi_i| - \rho_\mathbb{X}\right)^2,
\end{equation}
which can be evaluated as
\begin{equation}
\begin{aligned}
    &\int d\bm{\psi}~ \left(\rho_\mathbb{X}^{\rm est}-\rho_\mathbb{X}\right)^2\\
    =& \frac{1}{N^2}\sum_{i,j=1}^N \int d\psi_id\psi_j~ |\Psi_i\rangle\langle\Psi_i|\Psi_j\rangle\langle\Psi_j| - \frac{1}{N}\sum_{i}^N \int d\psi_i~ |\Psi_i\rangle\langle\Psi_i|\rho_\mathbb{X} - \frac{1}{N}\sum_{i}^N \int d\psi_i~ \rho|\Psi_i\rangle\langle\Psi_i| + \rho_\mathbb{X}^2\\
    =& \frac{1}{N^2}\sum_{i,j=1}^N \int d\psi_id\psi_j~ |\Psi_i\rangle\langle\Psi_i|\Psi_j\rangle\langle\Psi_j| -  \rho_\mathbb{X}^2 \\
    =& \frac{1}{N^2}\sum_{i=1}^N \int d\psi_i~ |\Psi_i\rangle\langle\Psi_i| + \frac{1}{N^2}\sum_{i\ne j}^N \int d\psi_id\psi_j~ |\Psi_i\rangle\langle\Psi_i|\Psi_j\rangle\langle\Psi_j| -  \rho_\mathbb{X}^2 \\
    =& \frac{1}{N} \left(1-\frac{1}{d_c} \right)\rho_\mathbb{X}.
\end{aligned}
\end{equation}
With this, the averaged Hilbert-Schmidt distance between $\rho_\mathbb{X}^{\rm est}$ and $\rho_\mathbb{X}$ can be bounded readily as
\begin{equation}
\begin{aligned}
    &\int d\bm{\psi}~ ||\rho_\mathbb{X}^{\rm est}-\rho_\mathbb{X}||_2\\
    = & \int d\psi_1\cdots d\psi_N~ \left[{\rm tr}\left(\rho_\mathbb{X}^{\rm est}-\rho_\mathbb{X}\right)^2\right]^{1/2}\\
    \le& \left[ \int d\psi_1\cdots d\psi_N~{\rm tr}\left(\rho_\mathbb{X}^{\rm est}-\rho_\mathbb{X}\right)^2\right]^{1/2}\\
    =& \left[ {\rm tr} \int d\psi_1\cdots d\psi_N~\left(\rho_\mathbb{X}^{\rm est}-\rho_\mathbb{X}\right)^2\right]^{1/2}
    \le \sqrt{\frac{1}{N}}.
\end{aligned}
\end{equation}
Here, $||X||_2\equiv\sqrt{ {\rm tr} XX^\dag}$ is the Hilbert-Schmidt norm (Frobenius norm). In the third line we used the H\"older's inequality, that is, for any measurable functions $f$ and $g$ on a measure space $S$ with measure $\mu$,
\begin{equation}
    \int_S |fg|d\mu \le \left(\int_S |f|^p d\mu\right)^\frac{1}{p}\left(\int_S |g|^q d\mu\right)^\frac{1}{q},
\end{equation}
with $p,q \in [1,\infty]$ and $1/p + 1/q =1$. Here, in our case,
\begin{equation}
    f=\left[{\rm tr}(\rho_\mathbb{X}^{\rm est}-\rho_\mathbb{X})\right]^{1/2}, \quad g=1,\quad p=q=2.
\end{equation}

The variance of the the Hilbert-Schmidt distance can be bounded more directly as
\begin{equation}
\begin{aligned}
\sigma^2 \equiv& \int d\bm{\psi}~ ||\rho_\mathbb{X}^{\rm est}-\rho_\mathbb{X}||_2^2 - \left(\int d\bm{\psi}~ ||\rho_\mathbb{X}^{\rm est}-\rho_\mathbb{X}||_2\right)^2
\\
\le & \int d\psi_1\cdots d\psi_N~ ||\rho_\mathbb{X}^{\rm est}-\rho_\mathbb{X}||_2^2\\
    = & \int d\psi_1\cdots d\psi_N~ {\rm tr}\left(\rho_\mathbb{X}^{\rm est}-\rho_\mathbb{X}\right)^2
    \le \frac{1}{N}.
\end{aligned}
\end{equation}

\section{General channels}

\begin{figure*}[t!]
    \centering
    \includegraphics[width=\textwidth]{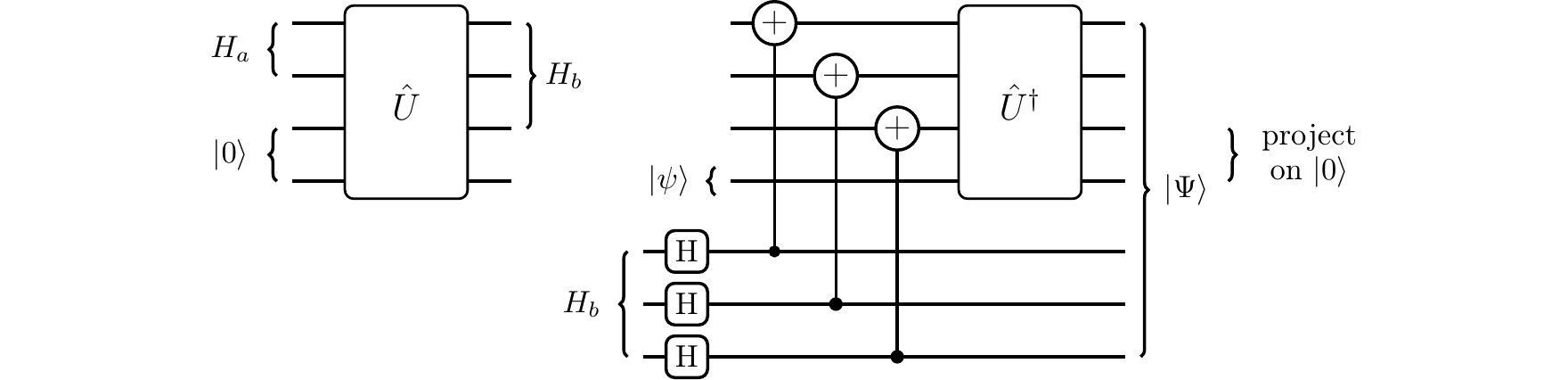}
    \caption{Left: The general channel $\mathbb{X}: L(H_a)\rightarrow L(H_b)$ can be dilated to a unitary channel $U$. Right: $U$ is used to generate random dual state $|\Psi\rangle$, which is then post-selected according to (\ref{eq:si_decomp}) to generate dual state $|\Phi\rangle$. The first moment of $|\Phi\rangle$ is the exact dual state of channel $\mathbb{X}$.}
    \label{fig:si_general}
\end{figure*}

A general channel $\mathbb{X}: H_a \rightarrow H_b$ can be dilated to a unitary channel $U$, such that for Hermitian operators $A$ and $B$ that applies on $H_a$ and $H_b$, respectively, 
\begin{equation}
   {\rm tr} \left[\mathbb{X}(A)B\right] = {\rm tr}\left[ U(A\otimes |0\rangle\langle 0|) U^\dag B\right],
\end{equation}
where $|0\rangle$ is a fixed state. This is illustrated in Fig. \ref{fig:si_general}~(left). Since $A\otimes |0\rangle\langle 0|$ has the same dimensionality as the unitary U, we can use the method developed in the previous sections to generate random states $|\Psi\rangle$ (Fig.~\ref{fig:si_general}, right), which form an exact duality relation
\begin{equation}
    {\rm tr} \left[\mathbb{X}(A)B\right] = d_U \cdot \int d\bm{\psi} \langle\Psi| A\otimes|0\rangle\langle0|\otimes B^t |\Psi\rangle,
\end{equation}
where $d_U$ is the dimension of $U$.

Let us decompose the random state $|\Psi\rangle$ as
\begin{equation}\label{eq:si_decomp}
    |\Psi\rangle = c_0 |0\rangle\otimes|\Phi\rangle + |\Tilde{\Psi}\rangle,
\end{equation}
where $|\Phi\rangle$ lives in the Hilbert space that support $A\otimes B$, and $|\Tilde{\Psi}\rangle$ is orthogonal to $|0\rangle$. $c_0^2=d_a/d_U\le 1$ is a normalization factor.
With this, the duality relation can be rewritten as
\begin{equation}
    {\rm tr} \left[\mathbb{X}(A)B\right] = d_a \cdot  \int d\bm{\psi} \langle\Phi| A\otimes B^t |\Phi\rangle.
\end{equation}
Therefore, our randomized duality relation is generalized to the generic channel $\mathbb{X}$, with the exact dual state
\begin{equation}
    \rho_{\mathbb{X}} \equiv \int d\psi |\Phi\rangle\langle\Phi|.
\end{equation}
Note that the duality relation of $\rho_{\mathbb{X}}$ implies that it is a normalized state. This is the reason why we chose the convention $c_0^2 = d_a/d_U$. $\rho_{\mathbb{X}}$ is also the transpose of the Choi matrix of $\mathbb{X}$. If we approximate $\rho_{\mathbb{X}}$ with $N$ realizations of $|\Phi\rangle$, 
    \begin{equation}
    \rho_{\mathbb{X}}^{\rm est} \equiv \frac{1}{N}\sum_k  |\Phi_k\rangle\langle\Phi_k|,
\end{equation}
the averaged Hilbert-Schmidt distance between $\rho_{\mathbb{X}}^{\rm est}$ and $\rho_{\mathbb{X}}$ can be bounded in the same manner as in the previous section, i.e., 
\begin{equation}
    \int d\bm{\psi}~ ||\rho_\mathbb{X}^{\rm est}-\rho_\mathbb{X}||_2
    \le \sqrt{\frac{1}{N}}.
\end{equation}

\section{Beyond the duality --- higher order correlations}
In the previous sections, we have established the exact channel-state duality, by evaluating the first moment of the random dual states for observables, i.e.,
\begin{equation}
     \mu_1 = \int d\psi~ \langle\Psi| A\otimes B |\Psi\rangle.
\end{equation}
In this section, we consider second order integral of the form
    \begin{equation}
     F = \int d\psi d\psi'~ |\langle\Psi| A\otimes B |\Psi\rangle'|^2.
\end{equation}
This quantity can be evaluate exactly as
\begin{equation}
\begin{aligned}
    F =& \int d\psi d\psi'~ {\rm tr}\left( A\otimes B |\Psi\rangle\langle\Psi|A\otimes B |\Psi\rangle'\langle\Psi'|\right)\\
    =& \int d\psi d\psi'~ {\rm tr}\left( UAU^\dag\otimes B |\psi\rangle \langle \psi|\otimes|\phi^+\rangle\langle\phi^+| UAU^\dag\otimes B |\psi'\rangle \langle \psi'|\otimes|\phi^+\rangle\langle\phi^+|\right)\\
        =& \int d\psi d\psi'~ {\rm tr}\left( UAU^\dag B^t \otimes I |\psi\rangle \langle \psi|\otimes|\phi^+\rangle\langle\phi^+| UAU^\dag B^t\otimes I |\psi'\rangle \langle \psi'|\otimes|\phi^+\rangle\langle\phi^+|\right).\\
\end{aligned}
\end{equation}
Now, the two integrals can be performed seperately as
\begin{equation}
    \int d\psi |\psi\rangle\langle \psi| = \frac{1}{d_c}I.
\end{equation}
Therefore,
\begin{equation}
\begin{aligned}
    F =\frac{1}{d^2_c} {\rm tr}\left( UAU^\dag B^t|\phi^+\rangle\langle\phi^+| UAU^\dag B^t |\phi^+\rangle\langle\phi^+|\right).
\end{aligned}
\end{equation}
Note that in the above expression, $UAU^\dag B^t$ only applies on one subsystem of the maximally entangled state $|\phi^+\rangle$. Let us decompose $|\phi^+\rangle$ as
\begin{equation}
    |\phi^+\rangle\langle\phi^+| = \frac{1}{d_b} \sum_{i,j}|i,i\rangle\langle j,j|.
\end{equation}
Then
\begin{equation}
\begin{aligned}
    F =& \frac{1}{d^2_a} \sum_{i,j}\sum_{m,n} {\rm tr}\left( UAU^\dag B^t|i,i\rangle\langle j,j| UAU^\dag B^t |m,m\rangle\langle n,n|\right)\\
    =&\frac{1}{d^2_a} \sum_{i,j} {\rm tr}\left( UAU^\dag B^t|i\rangle\langle j| UAU^\dag B^t |j\rangle\langle i|\right).\\
\end{aligned}
\end{equation}
Compared to the definition of the out-of-time order correlator (OTOC) for operators $W$ and $V$, with respect to a quantum state $\rho$
\begin{equation}
    F_\rho \equiv {\rm tr}\left(UWU^\dag V UWU^\dag V \rho\right),
\end{equation}
$F$ can be viewed as an averaged OTOC for operators $W=A$ and $V=B^t|i\rangle \langle j|$ evaluated with respect to a thermal state at infinite temperature. 

When $B=\Pi_B$ is a rank-$1$ projector in the computational basis---the basis of the maximally entangled state $|\phi^+\rangle$, $F$ simplies to a single OTOC, i.e.,
\begin{equation}
    F =\frac{1}{d^2_a} {\rm tr}\left( UAU^\dag \Pi_B UAU^\dag \Pi_B\right).\\
\end{equation}

\end{document}